\documentclass[10pt, conference, twocolumn, compsocconf]{IEEEtran}
\usepackage{supertabular}
\usepackage{array}

\usepackage{epsfig}
\usepackage{graphicx,xspace,xcolor,cite,url}
\usepackage{times}
\usepackage{subfigure}
\usepackage{algorithmic,algorithm}
\usepackage{xspace}
\usepackage{amssymb}
\usepackage{multirow}
\usepackage{booktabs}
\usepackage{listings}
\usepackage{comment}
\usepackage{fancybox}
\usepackage{xspace}
\usepackage{paralist}
\usepackage{multirow}

\usepackage[normalem]{ulem}
\newcommand{\etal}[0]{\emph{et al.\xspace}\xspace}

\IEEEoverridecommandlockouts
\begin{document}

\date{}

\title{\Large \bf New Problems and Solutions in IoT Security and Privacy}
\author{
  \IEEEauthorblockN{Earlence Fernandes}
  \IEEEauthorblockA{University of Wisconsin-Madison\\ earlence@cs.wisc.edu} \and
  
  \IEEEauthorblockN{Amir Rahmati}
  \IEEEauthorblockA{Stony Brook University\\ amir@cs.stonybrook.edu} \and

  \IEEEauthorblockN{Nick Feamster}
  \IEEEauthorblockA{University of Chicago\\ feamster@uchicago.edu}

}

\maketitle
\thispagestyle{plain}
\pagestyle{plain}

\begin{abstract}
In a previous article for S\&P magazine~\cite{fernandes2017internet}, we made a case for the new intellectual challenges in Internet of Things security research. In this article, we revisit our earlier observations, and discuss a few results from the computer security community that tackle new issues. Using this sampling of recent work, we identify a few broad general themes for future work. 
\end{abstract}

\section{Introduction}
\label{sec:intro}
Bruce Schneier, a noted computer security and cryptography expert, recently noted that ``Our cars are computers with an engine, our ATMs are computers with money inside, and our refrigerators are computers that keep things cold.'' The proliferation of the Internet of Things (IoT) is augmenting everyday objects with computational and networking capabilities. The IoT brings many benefits ranging from safety and security to efficiency. However, it also brings many security and privacy threats. In classical computer security, the maximal damage an attacker could inflict was limited to data loss. With the IoT, attackers can have physical effects in the world, such as opening doors~\cite{smartthings16}, causing fake fire alarms~\cite{smartthings16}, and disrupting electricity supply~\cite{soltan18}. The computer security and privacy community has recognized this emerging problem, and has focused its efforts on understanding and fixing its issues. 

An often-asked question is, are there any new intellectual challenges in securing these systems? This is a natural question to ask given that recent high-profile IoT security events like the Mirai attack~\cite{miraikrebs} largely depended on known traditional security issues (e.g., default passwords, lack of network security). We believe that like any emerging area, there are new problems, old ones in new forms, and problems we know how to solve. In our previous article, ``Internet of Things Security Research: A rehash of old ideas or new intellectual challenges?'', we made an early case for why there might exist intellectually novel and challenging problems in securing these systems. In this article, we revisit those observations, and discuss novel results that have arisen since then. As this is an active area, we take a broad view on IoT security and privacy, but we do not try to be complete in our treatment. Our goal is to provide a few pointers to recent work that brings new problems and solutions to light and, in doing so, identify a few general themes and future paths. 

\section{A sampling of recent results in IoT security}
\label{sec:overview}
In this article, we discuss a sampling of recent results that have appeared in academic security conferences, including work at conferences in related areas (Table~\ref{tab:papers}). We sampled papers to achieve broad coverage of novel themes---a comprehensive treatment of existing work is outside the scope of this article.  We organize our discussion based on the IoT computing stack introduced in our earlier article~\cite{fernandes2017internet}:

\begin{itemize}
    \item \textit{Hardware layer:} the embedded systems that make up IoT devices. These devices use sensors and actuators to sense and effect physical change in their environments. 
    \item \textit{Network layer:} the set of network protocols and other communication infrastructure that enable devices to talk with each other, and with other pieces of software that form an IoT ecosystem.
    \item \textit{Middleware layer:} the software and services that understands heterogeneous IoT protocols, and enables higher-level functionality to execute. This can include cloud-hosted services.
    \item \textit{Users and Applications layer:} the topmost layer that users can see. This layer is often specific to a domain. For example, in the context of smart homes, this layer could be if-this-then-that style programs.
\end{itemize}

The results we discuss in this article address issues in all these layers. Within a specific layer, we organize our discussion around a set of themes. The themes broadly characterize why the problems and solutions considered in the corresponding papers are novel.

\begin{table*}[t!]
    \centering
    \caption{Sampling of papers considered in this article, and some future challenges that arise as a result.}
    \label{tab:papers}
    \begin{tabular}{c c p{4cm} p{5cm}}
        \toprule
         \textbf{Category} & \textbf{Paper} & \textbf{General Themes in IoT Security} & \textbf{Potential Future Challenges}  \\
         \toprule
         Hardware, Physical Safety & Ding \etal~\cite{ding18} & \multirow{3}{4cm}{Properties of deployment environments that affect security guarantees; Physical-world aware security } & \multirow{3}{5cm}{Interplay between intermittently-powered devices and security;  Classic security mechanisms extended to be physically aware}  \\ 
         & Kiningham \etal~\cite{cesel} & & \\
         & Rahmati \etal~\cite{tardis} &  & \\
         & Simpson \etal~\cite{simpson2017securing} &  & \\
         & Soltan \etal~\cite{soltan18} &  & \\
         & & & \\ 
        
         \midrule
          
         Network & Formby \etal~\cite{formbyndss16} & \multirow{2}{4cm}{Behavior Predictability; Conflicting Security Properties; New Privacy Attacks and Defenses} & \multirow{2}{5cm}{Comprehensive investigations of conflicting properties and behavior predictability; Energy-security tradeoff for padding-based privacy defenses} \\
          & Wilson \etal~\cite{wilson} &  & \\
          & Apthorpe \etal~\cite{apthorpe} & & \\
          & & & \\
          \midrule
          
         Middleware, OS & Fernandes \etal~\cite{dtap} & \multirow{3}{4cm}{Decentralization of trust; Designing access control systems around environmental context; Global Safety and Security Properties} & \multirow{3}{5cm}{Exploring integrity, confidentiality, and availability problems for decentralized middleware; Designing access control systems with sensing as a core component; Model physical interactions } \\
          & Schuster \etal~\cite{schuster} & & \\
          & Celik \etal~\cite{celik} & & \\
          & & & \\
          & & & \\

          \midrule
          
         Users, Applications & He \etal~\cite{he} & \multirow{5}{4cm}{Interplay of usability, security, and multi-user devices; Attacks that rely on a lack of usability and on new interaction modalities; Risk asymmetry} & \multirow{5}{5cm}{Understanding the efficacy and usability of sensing for access control and authentication purposes; Understanding and mitigating inter-personal issues in multi-user devices; Defending against voice-based confusion attacks} \\
          & Kumar \etal~\cite{kumar} &  & \\
          & Rahmati \etal~\cite{tyche} & & \\
          & Zhang \etal~\cite{zhang} &  & \\
          & Zeng \etal~\cite{zeng} & & \\
          & & & \\
         
         \bottomrule
         
    \end{tabular}
    
\end{table*}

\subsection{Hardware Security and Physical Safety}



\noindent\textbf{Theme: Environmental Factors.} Embedded systems deployed in homes, buildings, cities, and on our bodies cannot be replaced as frequently as we replace our laptops or phones. For example, it is uncommon to replace a refrigerator every year, nor is it common to replace a pacemaker frequently. Furthermore, IoT devices and sensors may be deployed inside concrete, under water, or in other harsh and/or hard to reach conditions. The deployment environments of IoT devices lead to new challenges. One such challenge is to ensure security properties like encryption remain secure for long periods of time. Several crypto-systems have gone from being recommended to being insecure in less than 10 years~\cite{cesel}. Ensuring that a system remains secure for long periods of time is difficult because attackers adapt quickly. Kiningham \etal make this observation, and outline the CESEL architecture that can remain secure for 20 years~\cite{cesel}, a reasonable period of time for IoT devices. CESEL consists of five primitives specifically chosen to be useful in that time frame. These properties are based on a few fundamental primitives that are currently common to cryptographic systems and protocols such as nonces, and modular arithmetic. 
Another issue is how to patch these devices against new security vulnerabilities. This is especially challenging given that the IoT space is dominated by startups, many of whom will either perish to the market forces, ending support for their devices, or do not have the resources or technical knowledge to maintain an update infrastructure. Simpson~\etal~\cite{simpson2017securing} propose creation of a central security manager built on top of the smart home’s hub or gateway router to alleviate this problem in home IoT settings. However,  this solution only targets home IoT hubs and will not be applicable in settings where such a hub does not exist.
Harsh deployment conditions may also affect devices access to a constant source of energy and the IoT devices have to rely on energy harvesting from sources such as solar, vibration, or RF to satisfy their requirements.  Having intermittent access to energy poses new security challenges. Rahmati~\etal~\cite{tardis} for example, highlight the lack of trusted source of time in this context and propose coarse-grained time measurements using volatile memory decay. 

\noindent\textbf{Theme: Physical Attacks.} A standard attacker technique in computer networks is to move laterally between nodes---an attacker might compromise a specific user's computer, and then use that as a launch point to compromise other systems in the network. Ding \etal discussed how an attacker might use physical channels to achieve this lateral movement~\cite{ding18}. For instance, assume a home has a temperature sensor, a thermostat, and a user-created rule that automatically opens the windows if the temperature goes above a certain value. An enterprising thief might compromise the thermostat, and then realize that they can turn it on to increase temperature in the room, triggering the user-created rule to open the windows. Ding \etal also proposed techniques to detect such interaction chains automatically by taking into account physical interactions in addition to digital interactions. 

Similar to how botnets can disrupt network infrastructure and services, Attacks on the IoT can damage physical infrastructure. Stuxnet worm~\cite{stuxnet}, for example, targeted Iran's nuclear facilities and damaged its centrifuges. Similar attacks can potentially affect civilian infrastructure. Soltan \etal~\cite{soltan18} discuss attacks on the power grid using a set of compromised high-wattage consumer appliances like ovens, and heaters. Previous attacks on the power grid often involved vulnerabilities in the management interfaces or the SCADA systems that comprise the grid. However, the introduction of Internet-connected appliances violates the fundamental assumption in power grid design that the consumer patterns are not adversarial. 





\noindent\textbf{Future Challenges.} Stepping back, we remark on a few future challenges in hardware and physical layer security inspired by the papers discussed above. We expect that many IoT devices will be deployed in difficult-to-reach environments, such as sensors in concrete bridges that harvest power using vibrations, and sensors inside the human body. The properties of these environments introduce new challenges: (1) The interplay between security primitives and intermittently-powered or ultra-low powered devices requires further investigation. For example does a CESEL mote provide any properties that allow a sensor to pick up a security protocol from where it left off because it ran out of power? Developing cryptographic algorithms, secure hardware, or software targeting these applications is an open area of research.
(2) How do we extend current security mechanisms such as information flow control and model checking to account for physical-world interactions? (3) How can embedded systems be patched and secured against future attacks if they are no longer supported by their manufacturer?


\subsection{Network}


\noindent\textbf{Theme: Behavior Predictability.} Most general-purpose computer systems support varied functionality. This translates to highly varying network signatures, making security mechanisms like anomaly detection difficult due to errors. By contrast, IoT devices have well-defined functions---a door lock opens and closes, an industrial relay is either on or off. Thus, we observe that IoT device network behavior can be relatively stable compared to other kinds of computer systems. Formby \etal use this notion in their work on creating behavioral fingerprints for industrial control systems (ICS)~\cite{formbyndss16}. A key observation in their work is that ICS devices have simple and well-defined functions, leading to relatively stable network signatures. Stepping back, we see that a classic network security technique can leverage the unique aspects of IoT devices to reduce errors.


\noindent\textbf{Theme: Conflicting Properties.} The main job of the TLS protocol is to enable end-to-end encryption between two computer systems so that third parties cannot inspect communication. It is generally a good sign that IoT devices support TLS. However, many times users want to audit and inspect what these devices are saying about them and their activities. Unfortunately, the TLS protocol does not allow this kind of inspection, leading to a conflict between security and auditability. We do not run into this conflict in the traditional web setting because the user controls one end of the TLS connection. Wilson \etal made this observation and invented TLS-Rotate-and-Release (TLS-Rar). This protocol allows for trusted auditing devices to decrypt TLS streams without compromising future communication. 

\noindent\textbf{Theme: New Privacy Attacks and Defenses.} Apthorpe \etal recently showed that network observers can infer IoT device activity even when the streams are encrypted~\cite{apthorpe}. Such an inference in the general case is error-prone because computers are general purpose devices. However, the behavior predictability discussed above can enable high-fidelity reconstruction of device and user activity from encrypted streams. Investigating such attacks in the context of non-IP protocols is an interesting open area. Apthorpe \etal also present a set of defenses based on independent link padding showing, that it can be effective, and bandwidth-efficient.

\noindent\textbf{Future Challenges.} Stepping back, we remark on a few themes and open challenges in this space. Behavior predictability is a general emergent theme for IoT devices. This can enable new defenses, and new attacks as well. An open opportunity is to thoroughly investigate this issue along multiple dimensions such as non-IP protocols, and resource constraints. For example, Apthorpe \etal discussed a padding defense. An open question might be: What is the cost of such padding on energy-constrained devices that uses protocols like BLE? This introduces a new energy-security tradeoff into padding defenses. Another challenge concerns the theme of conflicting properties. Currently, it is unknown whether there are other security properties that we take for granted in the non-IoT world, but have undesirable properties in the IoT world. Investigating these conflicting properties in a more comprehensive manner is an open and novel research question.

\subsection{Middleware and OS}

\noindent\textbf{Theme: Decentralization of Trust.} A primary function of the middleware layer is to enable interoperability between multiple IoT ecosystems. Current consumer-grade IoT devices, such as those found in homes, and offices exhibit heterogeneity in their communication protocols and in their APIs. This heterogeneity exists due to resource constraints (e.g., Bluetooth Low Energy used in battery-powered presence detectors), a lack of standardization, and market competition. Middleware, such as Samsung SmartThings, and If-This-Then-That (IFTTT) understands many protocols, and enables programmers and users to create automations that span diverse sets of devices. Fernandes \etal examined the security design issues in the IFTTT platform~\cite{dtap}. The IFTTT platform uses OAuth tokens (a type of bearer credential) to gain access to a user's resources. Using these tokens, IFTTT can execute rules of the form: \texttt{if} trigger \texttt{then} action.  Fernandes \etal observed that middleware like IFTTT becomes a gold mine of tokens when we consider the scale at which these services operate. If such middleware is compromised, millions of security credentials will be leaked. Consequently, Fernandes \etal proposed \textit{Decentralized Action Integrity}, a security principle that provides the guarantee that even if the tokens are stolen, the attacker cannot use them arbitrarily. Although the notion of an integrity guarantee is not new, the IoT setting opens up new research opportunities for formulating and enforcing decentralized security properties. The general takeaway is that IoT systems are fundamentally decentralized, and security mechanisms should consequently be decentralized as well. Another novelty is that decentralization can address the `end-of-life' problem. Physical devices like door locks and refrigerators have long service lives, and can outlast the companies that make them. If an IoT device or middleware is dependent on infrastructure maintained by a company that could eventually stop a product line, will the automations continue to function? Worse yet, if a company goes out of business, who would maintain the infrastructure? Decentralization provides an answer by moving functionality closer to the end-nodes. DTAP exemplifies this idea by using the IFTTT-cloud only as compute infrastructure. Conceptually, any cloud service can replace the IFTTT-cloud component in DTAP architecture.


\noindent\textbf{Theme: Contextual Access Control.} Another function of middleware is to enforce proper access control and authorization. A defining characteristic of access control in the IoT setting is that many decisions are contextual. For example, access to a sprinkler might only be allowed when someone is at home. Schuster \etal explored the use of context in access control decisions, and observed that access control in the home is often decentralized in addition to being contextual, often occurring in multiple independent frameworks~\cite{schuster}. For example, the Nest thermostat and the LG oven might each implement their own context detection and access control framework. Instead of duplicating functionality, Schuster \etal propose the Environmental Situation Oracle (ESO) that encapsulates the sensing of physical contexts. In contrast to existing notions of context (e.g.,  what IP address range did the login occur from?), the novel challenge in the IoT setting is to sense physical phenomena (e.g., someone is near the oven) in a usable, scalable, and secure way.

\noindent\textbf{Theme: Global Interactions.} The middleware enables many different kinds of automations to run concurrently. A set of programs interacts with many physical devices in a single deployment. Even if individual programs are safe, unsafe situations can arise when we consider the behavior of this collection of programs as a whole. This is a very complex problem in the general case, making reasoning about global safety properties difficult. However, in the smart home setting, the individual programs are generally simple. Thus, examining global safety and security properties in the context of home automation programs is a promising and novel area of investigation. Recently, Celik \etal built the Soteria system that uses model checking to jointly investigate the security of IoT environments---a collection of physical devices and multiple automation programs~\cite{celik}.


\noindent\textbf{Future Challenges.} Stepping back, we remark on a few open challenges in this space. Like many computing systems before, we see IoT middleware architecture placing undue trust in centralized untrustworthy components. As many IoT systems are fundamentally decentralized, it is sensible to express security guarantees in a decentralized way as well. Specifically,  the challenge is to de-privilege centers of trust, such as monolithic web apps (e.g., IFTTT), into smaller units where each user only trusts a minimal amount of infrastructure. Additionally,  IoT deployments consist of a set of automations, and analyzing the security and safety of the environment as a whole is an open problem when we consider physical interactions between entities. Finally, another challenge concerns that fact that access control for IoT is physically contextual. Designing systems to sense and enforce contextual-policies presents new opportunities. Sensing of context in particular has strong connections to usability, and we explore this in the next section. 


\subsection{Users and Applications}
\label{sec:user}

\noindent\textbf{Theme: Usability of Contextual Authentication and Access Control.} Authentication in traditional computing systems generally involves mechanisms like user names and passwords. In a smart home setting, many devices do not have traditional input-output modalities such as a screen or keyboard. Thus, exploring new forms of establishing identity is an interesting new challenge. Biometrics is a popular alternative, but unlike passwords, it is inexact and can make errors. Understanding the impact of these errors, and corresponding user behavior is an open question. He \etal recently conducted a study to explore access control and authentication of \textit{users to devices}, and concluded that contextual factors and interpersonal relationships should play a more significant role in access control policies than they currently do~\cite{he}. Zeng \etal also report on a study investigating usability issues of current access control mechanisms in homes with multiple users, concluding that multi-user issues will be a new challenge~\cite{zeng}.

\noindent\textbf{Theme: Risk Asymmetry.} Another aspect of usability concerns how to present app permission requests to users. Current designs group device operations into similar functional units. For example, \texttt{oven.on} and \texttt{oven.off} are in the same functional group. However, physical device operations have clear and intuitive notions of risk: \texttt{oven.on} is a potential fire hazard, \texttt{oven.off} is potentially uncooked food. Rahmati \etal observe this risk \textit{asymmetry} and contributed a design process to leverage risk asymmetry as an alternative to functional groupings of permissions. The resulting system, Tyche, showed that apps can remain functional with 60\% less access to high-risk operations.


\noindent\textbf{Theme: Novel Attacks due to New Interaction Modalities.} Usability issues also permit novel attacks. Kumar \etal and Zhang \etal recently discovered voice-based confusion attacks on smart speakers like Amazon Alexa and Google Home~\cite{kumar,zhang}. An attack is a set of spoken words that sound similar, such as ``Hey Alexa, ask Capital One to...'' and ``Hey Alexa, ask Capitol Won to...'' These attacks leverage a combination of lack of user awareness about the specific application they are talking to, and the deficiencies in current speech processing technology. This is complicated by the fact that many voice assistant currently lack the general context humans have when conversing. Exploring methods to increase this context in the hope of disambiguating confusing situations is an open research area. 

\noindent\textbf{Future Challenges.} Stepping back, we discuss a few open challenges in this layer: (1) we realize that access control and authentication in homes is contextual, and will occur through non-traditional interaction modalities. Characterizing this space and building \textit{usable} systems that enforce the breadth of access control and authentication policies will probably require leveraging results from the ubicomp community in sensing; (2) characterizing and navigating the interpersonal issues in multi-user device situations presents new opportunities in usability security; (3) understanding the effectiveness of voice-based confusion attacks, and exploring defenses using biometrics and user education presents opportunities in enhancing the design of voice assistants.

\section{Concluding Remarks}
\label{sec:conclusion}
Using a sampling of recent results, we revisited our earlier S\&P article, and discussed intellectually novel and challenging problems in IoT security and privacy. We also identified general themes that are inspired by these results, and discussed a few open areas for future work at all layers of the computing stack. Although our treatment of results is incomplete, it is broad in its coverage of novel themes. We hope that this article spurs on research in IoT security and privacy in these new and challenging areas, so that we may gain the benefits of a connected world without the risks.

\newpage
\bibliographystyle{IEEEtranS}
\bibliography{references}

\end{document}